\documentclass[%
 aip,
 amsmath,amssymb,
 reprint,
]{revtex4-1}

\usepackage{graphicx}
\usepackage{dcolumn}
\usepackage{bm}

\usepackage[utf8]{inputenc}
\usepackage[T1]{fontenc}
\usepackage{mathptmx}
\usepackage{etoolbox}
\usepackage{color}
\usepackage[dvipsnames]{xcolor}
\usepackage{stackengine}
\usepackage{subfigure}
\usepackage{siunitx} 
\usepackage[nameinlink,capitalize]{cleveref}
\usepackage{placeins}

\newcommand{\newmk}[1]{{\color{black}#1}}
\newcommand{\new}[1]{{\color{black}#1}}
\newcommand{\newnew}[1]{{\color{black}#1}}
\makeatletter
\def\@email#1#2{%
 \endgroup
 \patchcmd{\titleblock@produce}
  {\frontmatter@RRAPformat}
  {\frontmatter@RRAPformat{\produce@RRAP{*#1\href{mailto:#2}{#2}}}\frontmatter@RRAPformat}
  {}{}
}%
\makeatother
\begin{document}

\preprint{}

\title[Numerical investigation of the segregation of emulsions]
{Numerical investigation of the segregation of turbulent emulsions}

\author{T. Trummler}
\author{A. Begemann}
\author{E. Trautner}
\author{M. Klein}

\email{theresa.trummler@unibw.de}
 \affiliation{
  Institute of Applied Mathematics and Scientific Computing, Bundeswehr University Munich\\ 
   Werner-Heisenberg-Weg 39, 85577 Neubiberg, Germany 
}
 
\date{\today}

\begin{abstract}
    We study the segregation of emulsions in decaying turbulence using direct numerical simulations (DNS) in combination with the volume of fluid method (VOF). To this end, we generate emulsions in forced homogeneous isotropic turbulence and then turn the forcing off and activate gravitational acceleration. This allows us to study the segregation process in decaying turbulence and under gravity.
    
    We consider non-iso-density emulsions, where the dispersed phase is the lighter one. The segregation process is driven by both the minimization of the potential energy achieved by the sinking of the heavier phase, as well as the minimization of the surface energy achieved by coalescence. To study these two processes and their impacts on the segregation progress in detail, we \newmk{consider different \new{buoyancy forces} and surface tension coefficients in our investigation, resulting in five different configurations.} The surface tension coefficient also alters the droplet size distribution of the emulsion.

    Using the three-dimensional simulation results and the monitored data, we analyze the driving mechanisms and their impact on the segregation progress in detail. We propose a dimensionless number that reflects the energy release dominating the segregation. Moreover, we evaluate the time required for the rise of the lighter phase and study correlations with the varied parameters gravitational acceleration and surface tension coefficient.
\end{abstract}

\maketitle

\section{Introduction}
  \label{s:Introduction}

  Emulsions are suspensions of immiscible liquids (such as oil and water) and play a central role in a wide range of industrial processes such as food processing~\citep{walker2015development, zhang2015influence,guzey2006formation}, pharmaceutical processes~\citep{spernath2006microemulsions} or oil production~\citep{dicharry2006stability,angardi2021critical}. 
  Moreover, current research is examining the application of fuel--water emulsions for more efficient and environmentally friendly power generation. Examples include gasoline--water direct injection (GWDI) for future gasoline engines~\citep{hoppe2017evaluation,heinrich2017gasoline} or fuel--water emulsions for small gas turbines~\citep{chmielewski2020combustion} and diesel engines~\citep{welscher2021comprehensive}. In particular for power generation applications, a better understanding of the stability of emulsions and the timescale of the segregation process is of central importance. To this end, we numerically study the segregation of emulsions in decaying turbulence under gravity.
 
  The formation of an emulsion requires energy input in form of kinetic energy, to deform and break up droplets. For a non-iso-density emulsion, the mixing of the lighter and the heavier phase requires additional energy input. In order to keep an emulsion stable, a continuous supply of energy is then required. Without further energy input, emulsions are unstable due to the natural tendency to minimize the potential and surface energy. To minimize the net potential energy, the heavier phase sinks, which is governed by the gravitational acceleration $g$ and the density difference between the two phases. To minimize the surface energy, droplets coalesce, thus reducing the interface area. A higher surface tension $\sigma$ leads to a higher variation of surface energy and thus increases the tendency for coalescence. Furthermore, it should be noted that interface minimization can be prevented by the presence of surfactants~\citep{kilpatrick2012water,goodarzi2019comprehensive}, such as the naturally occurring surfactants asphaltene and resins. Both rising and coalescence lead to segregation of emulsions, which can be quantified by the height of the lighter phase, i.e., the position of its center of mass in the direction of the gravitational acceleration, and the interface area, respectively. However, the two processes of rising and coalescence mutually interact with each other since larger droplets rise more easily and acceleration in one direction promotes coalescence.

  Experimental studies on the segregation process of emulsions mainly focus on chemical engineering aspects and monitor the height of the coalescing interface. There are several studies in the literature related to modeling of gravity assisted oil-water emulsion separation in oil production processes~\citep{dalingaros1987prediction, jeelani1998effect,jeelani1985prediction,henschke2002determination, jeelani2005creaming, lobo1993dispersion,aleem2020experimental}. The proposed models are designed for gravity settlers to separate the water during the oil production process and deliver correlations for the temporal evolution of zone heights. A review of these models is provided by \citet{frising2006liquid}. Most of these models require various input and modeling parameters and are very sophisticated. Further, limited optical access makes experimental studies of emulsion segregation processes challenging and requires advanced measurement techniques~\citep{koegl2020characterization}. For this reason, information on the interface area in segregating emulsions is hard to access. In this work, we want to complement these experimental studies with a numerical investigation for a generic configuration. 

  Several numerical studies of emulsions and emulsification processes have been reported in the past, mostly focusing on droplet size distributions. First numerical emulsion studies employed Lattice-Boltzmann (LB) methods, such as e.g. \citet{Perlekar:2012ip,Skartlien:2013ec} and \citet{Mukherjee:2019ka}. More recently, \citet{CrialesiEsposito:2021ui} and \citet{Begemann:2022CJ} utilized direct numerical simulations (DNS) combined with the volume of fluid method (VOF) for these investigations. Moreover, several numerical studies focused specifically on the break-up of droplets such as e.g. \citet{Komrakova:2019kz} or \citet{Shao:2018gg}. The reverse process to emulsification, namely coalescence and rising of the lighter phase, was studied in the following papers: \citet{Dodd:2016eo} investigated droplet coalescence and droplet turbulence interaction in decaying turbulence and found that the energy release due to coalescence processes has an impact on the decay of the turbulent kinetic energy. 

  \new{An important and central aspect of multiphase configurations is the effect of the buoyancy force due to a density difference between the phases and gravity. Previous numerical studies on the effect of the buoyancy force considered, for example, bubble-laden downflow configurations~\citep{brauer2021turbulent,Trautner:2021de,hasslberger2020direct} or rising bubbles~\citep{meller2022sub,hasslberger2018flow}. In addition, \citet{Saeedipour:2021gg} and \citet{estivalezes2022phase} have recently performed simulations of the phase inversion test case, where the lighter phase is initialized at the bottom of a box and rises due to gravitational acceleration. Despite the central importance of the buoyancy force on the segregation of non-iso-density emulsions, we are not aware of any previous numerical simulation studies on this. The effect of the buoyancy force on emulsion stability and emulsion segregation is of particular importance for various applications, especially with respect to emulsions in power generation (see above). With the present work, we aim to complement experimental studies~\citep{al2009experimental, sazonov2019untersuchung} with numerical investigations of emulsion stability and segregation. CFD simulations can overcome some of the limitations and challenges in experimentally characterizing emulsions~\citep{mcclements2007critical,koegl2020characterization} and provide new and more detailed insights.}

  The present work builds upon our recent paper on emulsification and emulsions~\citep{Begemann:2022CJ}. \new{Using the enhanced linear forcing approach proposed in our recent paper, we can generate a statistically stationary emulsion with a prescribed turbulent kinetic energy, and therewith obtain well-defined initial conditions for studying the segregation. In the present work, we} study the segregation of emulsions resembling oil--in--water \new{liquid--liquid} emulsions in terms of density ratio. In our study, we vary \new{the buoyancy force (by varying the gravitational acceleration $g$)} and the surface tension coefficient $\sigma$, the latter resulting in different droplet size distributions of the emulsions. Hence, we focus on the parameters affecting the segregation progress (minimization of potential and surface energy). For our studies, we use DNS with the finite volume approach and the VOF method. The emulsions are generated by a linear forcing of turbulence augmented with a PID controller~\citep{Begemann:2022CJ}. We then switch off the forcing and activate the gravitational acceleration and let the emulsions segregate in decaying turbulence under gravity. 

  The paper is structured as follows. In \cref{s:Method}, we describe the computational method. \Cref{s:Configs} presents the considered configurations and the numerical setup. The results are presented in \cref{s:Results}, which firstly studies the segregation process in detail, then focuses on the energy releases to elucidate the dominant mechanisms and finally studies the timescale of the segregation. \Cref{s:Conclusions} summarizes the findings and draws conclusions.

\section{Computational method}
 \label{s:Method}

  The simulations are conducted with the open source code PARIS (PArallel, Robust, Interface Simulator)~\citep{Aniszewski:2021cpc}. PARIS has been specifically designed for simulations of multiphase flows and is often used for studies of atomization processes, see e.g.~\citep{klein2017comparison,hasslberger2019flow,salvador2018analysis,ling2017spray} as well as other multiphase flow configurations~\citep{hasslberger2018flow}.
  
  The solver uses the single fluid formulation \citep{ProsperettiTryggvason:2007CompMeth} of the incompressible Navier-Stokes equations. The continuity and momentum equation are given as
  \begin{equation}
    \label{eq:continuity}
    \frac{ \partial u_{ i } }{ \partial x_{ i } } = 0\,,
  \end{equation}
  \begin{equation}
    \label{eq:momentum}
    \begin{split}
      & \rho \left( \frac{ \partial u_{ i } }{ \partial t } + \frac{ \partial u_{ i } u_{ j } }{ \partial x_{ j } } \right) = \\
      & - \frac{ \partial p }{ \partial x_{ i } } + \frac{ \partial }{ \partial x_{ j } } \left[ \mu \left( \frac{ \partial u_{ i } }{ \partial x_{ j } } + \frac{ \partial u_{ j } }{ \partial x_{ i } } \right) \right] + \sigma n_{ i } \kappa \delta_{ s } + \rho g_{ i }
    \end{split}
  \end{equation}
  with the density $\rho$, the dynamic viscosity $\mu$, the $i^{th}$ velocity component $u_i$, the pressure $p$ and the gravitational acceleration $g_i$. In each cell, the density and viscosity values are linearly interpolated using the local volume fraction $\alpha$ of the dispersed phase, which is tracked with the geometrical VOF method~\citep{HirtNichols:1981jcp}:
    \begin{equation}
       \label{eq:densityViscosity}
       \rho=\alpha \rho_d + \left( 1-\alpha \right) \rho_c,\quad \mu=\alpha \mu_d + \left( 1-\alpha \right) \mu_c.
    \end{equation}
  The subscripts $d$ and $c$ denote the dispersed and the carrier phase, respectively.
  
  The Continuous-Surface-Force (CSF) approach \citep{BrackbillKotheZemach:1992jcp} determines the surface tension force from the surface tension coefficient $\sigma$, the interface normal $n_i=\frac{\partial \alpha}{\partial x_i}/| \nabla \alpha |$, the interface indicator function $\delta_S=|\nabla \alpha|$ and the interface curvature $\kappa=\frac{\partial n_i}{\partial x_i}$. The latter is accurately computed using a state-of-the-art height function approach~\citep{Popinet:2009jcp}. Details on its implementation in PARIS can be found in \citet{Aniszewski:2021cpc}.
  
  The advection of the VOF marker function is performed using a geometrical interface reconstruction algorithm. The respective transport equation is given as
    \begin{equation}
      \label{eq:VOFtransport}
      \frac{\partial \alpha}{\partial t}+u_i \frac{\partial \alpha}{\partial x_i}=0,\,\,\,\alpha=
      \begin{cases}
      1, & \text{if $\pmb{x}$ is in dispersed phase.}\\
      0, & \text{if $\pmb{x}$ is in carrier phase.}
      \end{cases}
    \end{equation}
  A red-black Gauss-Seidel solver with overrelaxation is employed to solve the Poisson equation for pressure in the framework of the projection method. The simulation is advanced in time using a second-order predictor-corrector method. For the spatial discretization, the finite-volume approach is realized using a cubic grid. The velocity components are stored on a staggered grid, while the pressure and the VOF marker function, as well as the local densities and the viscosities resulting from the latter, are computed at the cell centers. The third-order Quadratic Upstream Interpolation for Convective Kinematics (QUICK) scheme~\citep{leonard1979stable} has been chosen to discretize the convective term of the momentum equation, while its viscous term is treated using central differences \newmk{and therewith second-order accurate.}

  \begin{figure*}[!htb]
  \centering
    \subfigure[]{\includegraphics[width=0.23\linewidth]{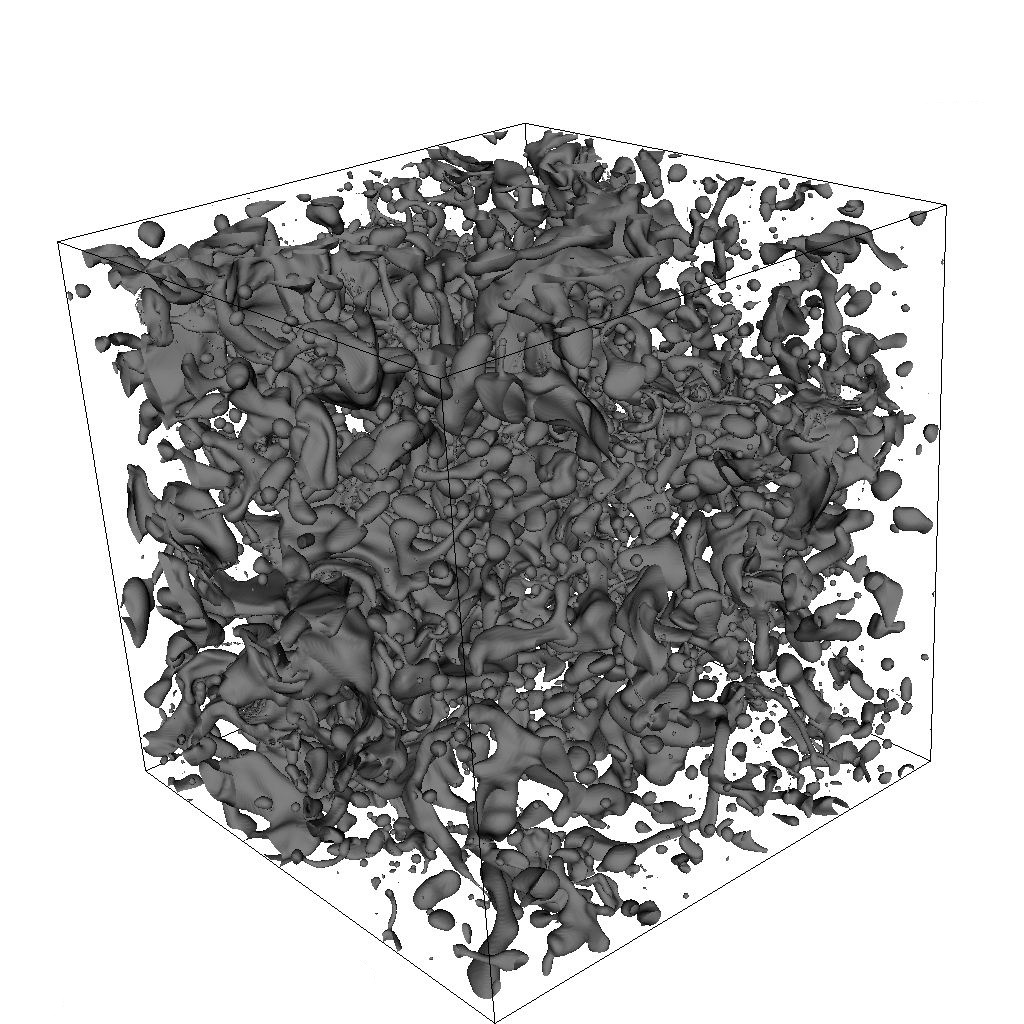}}
    \subfigure[]{\includegraphics[width=0.23\linewidth]{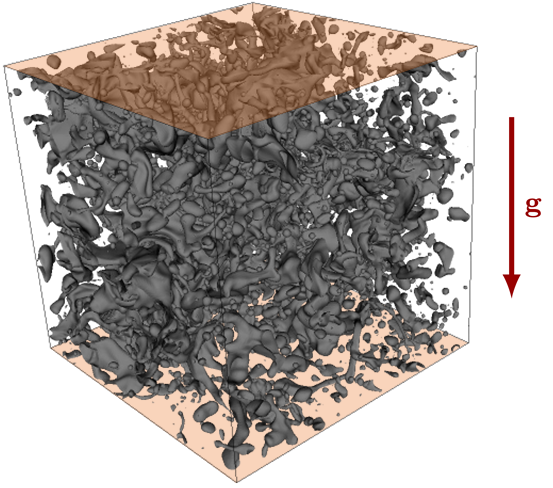}}
    \hspace{0.2cm}
    \subfigure[]{\includegraphics[width=0.22\linewidth]{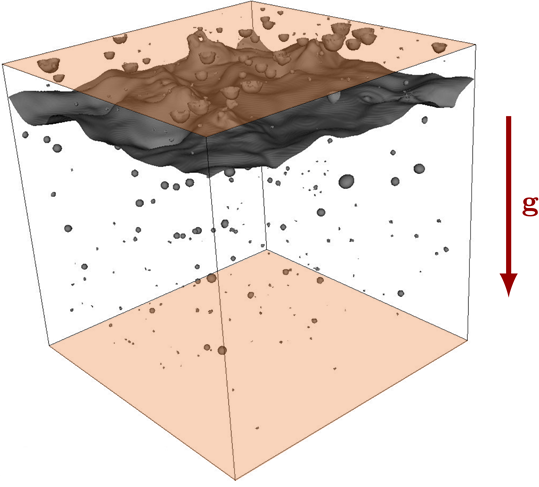}}
   \caption{
     Simulation setup. (a)~Turbulent emulsion at statistically stationary state, (b)~forcing is turned off, gravitational acceleration $g$ is activated and slip walls are prescribed in the direction of the gravitational acceleration (orange walls), (c)~segregation under $g$. 
   }
   \label{fig:sim_setup}
  \end{figure*}%
  
  To generate a turbulent emulsion in homogeneous isotropic turbulence (HIT), we employ the linear Lundgren forcing~\citep{Lundgren:2003uf} extended by a PID controller, see \citet{Begemann:2022CJ}. This extension provides a constant turbulent kinetic energy and accelerates the emulsification process.

  \begin{table}[!tb]
    \center
    \caption{Constant emulsion parameters.}
    \label{tab:Setup} 
    \begin{tabular}{ccccccccc} 
      \hline
      \multicolumn{1}{c}{$\phi$} & 
      \multicolumn{1}{c}{$\rho_c$} & 
      \multicolumn{1}{c}{$\rho_d$} &
      \multicolumn{1}{c}{$\nu_c=\nu_d $} & 
      \multicolumn{1}{c}{$k$ } & 
      \multicolumn{1}{c}{$\varepsilon$ } & 
      \multicolumn{1}{c}{$Re_\lambda$}&
      \multicolumn{1}{c}{L} & 
      \multicolumn{1}{c}{N}  \\
      - & $\si{kg/m^3}$&$\si{kg/m^3}$& $\si{m^2/s}$&$\si{m^2/s^2}$ &$\si{m^2/s^3}$& -&$\si{m}$& - \\
      \hline  
       1/8 & 1 & 0.9 & 0.001 & 0.5 & 0.153 &104& $2\pi$ & 384\\
      \hline
    \end{tabular}
  \end{table}

  \begin{table}[!tb]
    \center
    \caption{Considered cases. The baseline values \newmk{(BL)} are $\sigma_\mathrm{BL} = \num{2e-2}\,\si{N/m}$, $d_{H\mathrm{BL}} = 0.1468\,\si{m}$ and $g_\mathrm{BL}=4.59\,\si{m/s^2}$. Note that the correlation between $d_{H}$ and $\sigma$ for constant $\rho_c$ and $\varepsilon$ reads $d_{H} = \sigma^{3/5}$, see also \cref{eq:Hinze}. The last three columns contain the dimensionless segregation number $Seg$, the dimensionless energy release ratio $\Psi$ and the ratio $Seg/\Psi$, see \cref{ss:Mechanisms} for details.}
    \label{tab:CasesSeg} 
    \begin{tabular}{lrrrrrrr} 
      \hline
      \multicolumn{1}{l}{Case} & 
      \multicolumn{1}{c}{\newnew{$We_l$}} & 
      \multicolumn{1}{c}{$\sigma/\sigma_\mathrm{BL}$} & 
      \multicolumn{1}{c}{$d_H/d_{H\mathrm{BL}}$} & 
      \multicolumn{1}{c}{$g/g_\mathrm{BL}$}& 
      \multicolumn{1}{r}{$\;\;\;Seg\;$} &      
      \multicolumn{1}{c}{$\;\;\;\Psi\;$}&      
      \multicolumn{1}{c}{$Seg/\Psi$}\\
      \hline
      \textit{Baseline}      & \newnew{21} & 1.0 & 1.00 & 1.0 & 10.6 & 2.5 & 4.2\\
      \textit{Low}~$g$       & \newnew{21}  & 1.0 & 1.00 & 0.5 & 5.3  & 1.3 & 4.2\\  
      \textit{High}~$g$      & \newnew{21} & 1.0 & 1.00 & 2.0 & 21.2 & 5.0 & 4.2\\  
      \textit{Low}~$\sigma$  & \newnew{70}  & 0.3 & 0.49 & 1.0 & 18.3 & 4.6 & 4.0\\ 
      \textit{High}~$\sigma$ & \newnew{4}  & 5.0 & 2.63 & 1.0 & 5.6  & 1.8 & 3.2\\       
      \hline
    \end{tabular}
  \end{table}

\section{Considered configurations}
  \label{s:Configs}

  In this paper, we study the segregation of emulsions under \new{different buoyancy forces obtained by varying gravitational acceleration $g$}. Further, we also consider emulsions with different droplet size distributions, which are obtained by a variation of the surface tension coefficient $\sigma$. 

  Emulsions feature polydisperse droplet size distributions. A reference value for droplet size distributions in emulsions is given by the Hinze scale $d_H$~\citep{hinze1955fundamentals}, which is expected to be the most stable maximum droplet diameter in emulsions for HIT. $d_H$ is determined by  
    \begin{equation}
      d_H= \left(We_{d,crit}/2 \right)^{3/5} \left( \rho_c/\sigma \right)^{-3/5}\varepsilon^{-2/5} \,, 
      \label{eq:Hinze}
    \end{equation}
  where $We_{d,crit}$ denotes the critical \newnew{droplet} Weber number, for which we assume $We_{d,crit}= 1.17$ following \citet{hinze1955fundamentals} and recent numerical studies~\citep{CrialesiEsposito:2021ui, Mukherjee:2019ka}. Moreover, $\rho_c$ is the density of the carrier fluid, $\sigma$ the surface tension coefficient and $\varepsilon$ the dissipation rate associated with the turbulence intensity.

    \newnew{For emulsions, a Weber number $We_l$ using a characteristic length scale can be defined as 
        \begin{equation}
            We_{l} = \frac{\rho_c {u^\prime}^2 l}{\sigma},  
            \label{eq:Wel}
        \end{equation}%
  taking into account the effect of surface tension. In experimental studies (e.g.~\citep{Perlekar:2012ip}), for example, the diameter of the stirrer is used as length scale. For the HIT considered here, the integral length scale of the turbulent flow field is employed. Therefore, emulsions at stationary state can be characterized by $We_{l}$ as demonstrated and discussed in \citet{Begemann:2022CJ}. Moreover, using $l = ({u^{\prime}}^2)^{3/2}/\varepsilon$,
  the correlation 
    \begin{equation}
    d_H/l \propto We_{l}^{-3/5} \,   
    \label{eq:dHWel} 
  \end{equation}%
  is obtained.
  }

  The emulsions are generated in HIT with a constant turbulent kinetic energy $k$. For isotropic turbulence, $k$ is given by $k = (3/2) {u^{\prime}}^2$, where $u^\prime$ denotes the velocity fluctuation. In case of linear forcing, the dissipation rate $\varepsilon$ is determined by $k$ and the integral length scale $l$, which is $20\%$ of the domain length~\citep{Rosales:2005kp,Carroll:2013jt}, using the correlation $l = ({u^{\prime}}^2)^{3/2}/\varepsilon$. Additionally, for HIT, a characteristic dimensionless number is the Taylor Reynolds number $Re_\lambda = \lambda u^\prime/\nu$ formed with the Taylor micro-scale $\lambda = \sqrt{15\nu/\varepsilon}u^\prime$. 

  We simulate five different configurations. \newmk{The parameters common for all configurations} are summarized in \cref{tab:Setup}. The volume fraction of the dispersed phase is $\phi = V_d/(V_d + V_c) = 1/8 $ for all cases. The carrier and dispersed fluid have a density of $\rho_c=1\,\si{kg/m^3}$ and $\rho_d=0.9\,\si{kg/m^3}$, respectively, thus making the dispersed fluid the lighter one. The kinematic viscosities are both set to $\nu_d=\nu_c=0.001\,\si{m^2/s}$. We here study emulsions generated at a turbulence intensity of $k=0.5\,\si{m^2/s^2}$ in a cubic domain with length $L = 2\,\pi$ and thus a Taylor Reynolds number of $Re_\lambda = 104$ and a dissipation rate of $\varepsilon=0.153\,\si{m^2/s^3}$. We discretize the domain with $N=384$ cells in each direction ($\approx \num{57e6}$ cells in total). This grid resolution has been chosen to fulfill the criterion $K_{max}\eta\geq 1.5$, see e.g.~\citet{Pope:2001turb}, where $K_{max}$ is the maximum wavenumber $K_{max}=N \pi /L$ and $\eta$ is the Kolmogorov scale $\eta=(\nu^3/\varepsilon)^{1/4}$, given by the kinematic viscosity $\nu$ and the dissipation rate $\varepsilon$. A grid study of this configuration can be found in our recent paper~\citep{Begemann:2022CJ}. 

  For the simulation of the segregation process, first, turbulent emulsions in HIT are generated as described in detail in \citet{Begemann:2022CJ}.  \new{We consider a cubic box with periodic boundary conditions.} \new{In order to generate turbulent emulsions,} we \new{first} perform single-phase simulations to obtain a fully developed single-phase HIT. Then we initialize  the dispersed phase as spherical droplets, which break up in the linearly forced HIT and an emulsion is obtained as visualized in \cref{fig:sim_setup}~(a). At a statistically stationary state, we turn the forcing off and prescribe slip walls in the direction of the gravitational force, see \cref{fig:sim_setup}~(b). A thin layer (four cells) of the carrier fluid is additionally initialized at the bottom to detach dispersed structures there. Then, we let the emulsions segregate (\cref{fig:sim_setup}~(c)). The time $t$ is measured from the time instant when the forcing is turned off. 

  \Cref{tab:CasesSeg} lists the considered configurations for the segregation. \newnew{The configurations are adopted from our previous investigation~\citep{Begemann:2022CJ} of emulsification processes and emulsions at statistically stationary state. The emulsions before segregation can be described by the dimensionless Weber number $We_l$.} Starting from a baseline case (\textit{Baseline}), the gravitational acceleration $g$ for the segregation process is varied. More precisely, $g$ is halved in case \textit{Low}~$g$ and doubled in case \textit{High}~$g$ with respect to the baseline case. Furthermore, also the surface tension coefficient $\sigma$ is varied for the cases \textit{Low}~$\sigma$ and \textit{High}~$\sigma$, resulting in different droplet size distributions at the statistically stationary state before the segregation. Note that for constant $\rho_c$ and $\varepsilon$, which is the case here, the correlation between $d_{H}$ and $\sigma$ reads $d_{H} = \sigma^{3/5}$, see also \cref{eq:Hinze}. For the variation of $\sigma$, the entire emulsification process has been simulated to obtain the respective emulsions at statistically stationary state. \newnew{The employed value for the surface tension $\sigma_\mathrm{BL} = \num{2e-2}\,\si{N/m}$ approximates that of realistic fluids~\citep{koegl2020characterization}. The gravitational acceleration $g_\mathrm{BL}=4.59\,\si{m/s^2}$ has been chosen to obtain a Bond number $Bo=\Delta \rho g R^2/\sigma$ of Hinze droplets comparable to realistic examples of liquid--liquid emulsions (estimated values $\Delta \rho = 100\,\si{kg/m^3}$, $g=9.81\,\si{m/s^2}$, $R=\num{1.6e-3}\,\si{m}$, $\sigma = \num{2e-2}\,\si{N/m}$). Additionally, it is worth noting that for gravity-driven rising/falling of a dispersed phase, the ratio of the density difference to the density of the disperse phase $\Delta \rho /\rho_d$ is decisive, and the ratio in our simulations corresponds to that of realistic  liquid--liquid emulsions.}
 
  To monitor the segregation progress, we track the interface area $A$ and the center of mass of each phase $h_i$. The interface area of the dispersed phase is numerically approximated with the volume integral of the gradient of the VOF marker function $|\nabla\alpha|$. As reference for the interface area, the theoretically completely segregated state with $A_{\infty} =L^2$ is used. 

  \begin{figure*}[]
  \centering
    \includegraphics[width=0.75\linewidth]{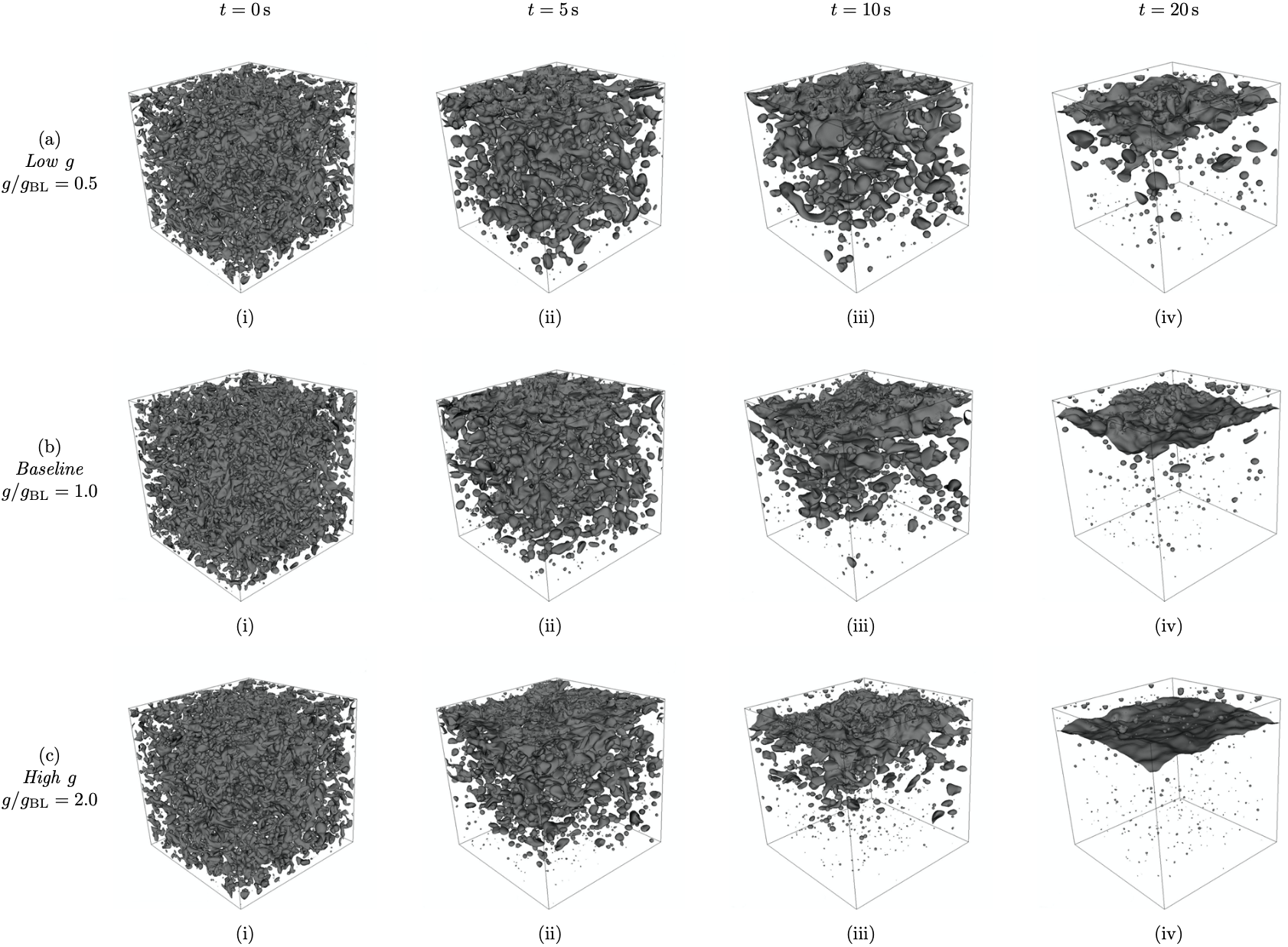}
       \caption{Visualization of the segregation process for varying $g$. Rows correspond to different $g$ values with (a) \textit{Low}~$g$, (b) \textit{Baseline}, (c) \textit{High}~$g$, and columns to different time instants (i, ii, iii, iv) ($t=\{ 0,\,5,\,10,\,20\,\si{s}\}$). The images show the iso-surface of the volume fraction corresponding to $\alpha = 0.5$.}
   \label{fig:SegG}
  \end{figure*}%

  \begin{figure*}[]
  \centering
    \includegraphics[width=0.75\linewidth]{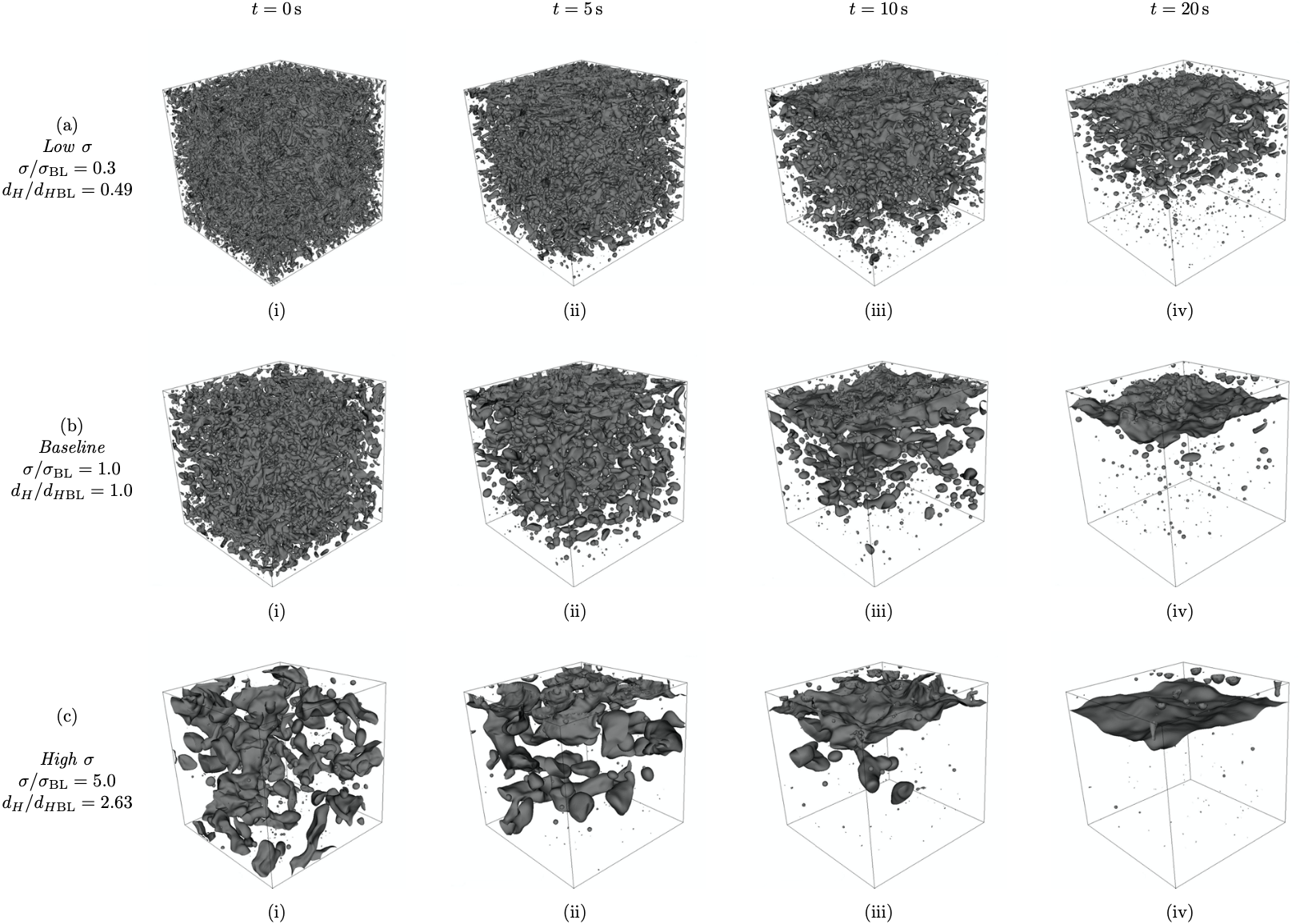}
   \caption{Visualization of the segregation process for varying $\sigma$. Rows correspond to different $\sigma$ values with (a) \newnew{\textit{Low}~$\sigma$}, (b) \textit{Baseline}, (c) \newnew{\textit{High}~$\sigma$}, and columns to different time instants (i, ii, iii, iv) ($t=\{ 0,\,5,\,10,\,20\,\si{s}\}$). The images show the iso-surface of the volume fraction corresponding to $\alpha = 0.5$.}
   \label{fig:SegS}
  \end{figure*}

  \section{Results}
  \label{s:Results} 

  We first study the segregation process for varying $g$ and $\sigma$ based on the obtained simulation results (\cref{ss:Results}). In \cref{ss:Mechanisms,ss:ERR}, we analyze the segregation process from the perspective of the energy releases driving it. At the end of this section (\cref{ss:u}), we derive correlations to estimate the characteristic velocity and timescale of the segregation process. 

  \subsection{Effect of $g$ and $\sigma$}
  \label{ss:Results}

  \Cref{fig:SegG} visualizes the segregation for different gravitational accelerations $g$, while \cref{fig:SegS} shows the same process for emulsions with different surface tension coefficients $\sigma$. In \cref{fig:SegG} (variation of $g$), a difference in the segregation progress can already be seen at the second time step visualized (see \cref{fig:SegG}~(ii)). At a higher $g$, a larger fraction of the lighter phase has risen and fewer dispersed structures are visible in the lower part of the box. As time progresses, the faster segregation at higher $g$ becomes more evident. Finally, at the last time step shown (see \cref{fig:SegG}~(iv)), a clear difference for different $g$ can be seen. At the highest $g$ (see \cref{fig:SegG}~(c, iv)) only a few very small structures are visible in the lower part. Further, it should be noted that in all configurations (a--d), some structures of the heavier phase are enclosed at the upper boarder, resulting in the interfaces visible there.  

  \Cref{fig:SegS} illustrates the segregation for different $\sigma$. Here, the distribution of the dispersed phase in the emulsion differs significantly. The higher the surface tension coefficient, the larger are the structures of dispersed fluid. For the case \textit{Low}~$\sigma$ (see \cref{fig:SegS}~(a)), there are many small structures with a smaller buoyancy force, which is proportional to $\Delta \rho V$, higher drag forces as well as more interactions in between dispersed structures. For these reasons, the segregation progress at lower $\sigma$ is significantly slower than for the cases with a higher $\sigma$. At the highest $\sigma$ (see \cref{fig:SegS}~(c)), comparably large structures are present and they experience a higher buoyancy force than smaller structures. Additionally, the high $\sigma$ promotes coalescence even more. For the \textit{High}~$\sigma$ case, a nearly complete segregation is reached at the last time step visualized (see \cref{fig:SegS}~(c,iv)).  

    \begin{figure*}[]
    \centering
    \subfigure{\includegraphics[width=0.99\linewidth]{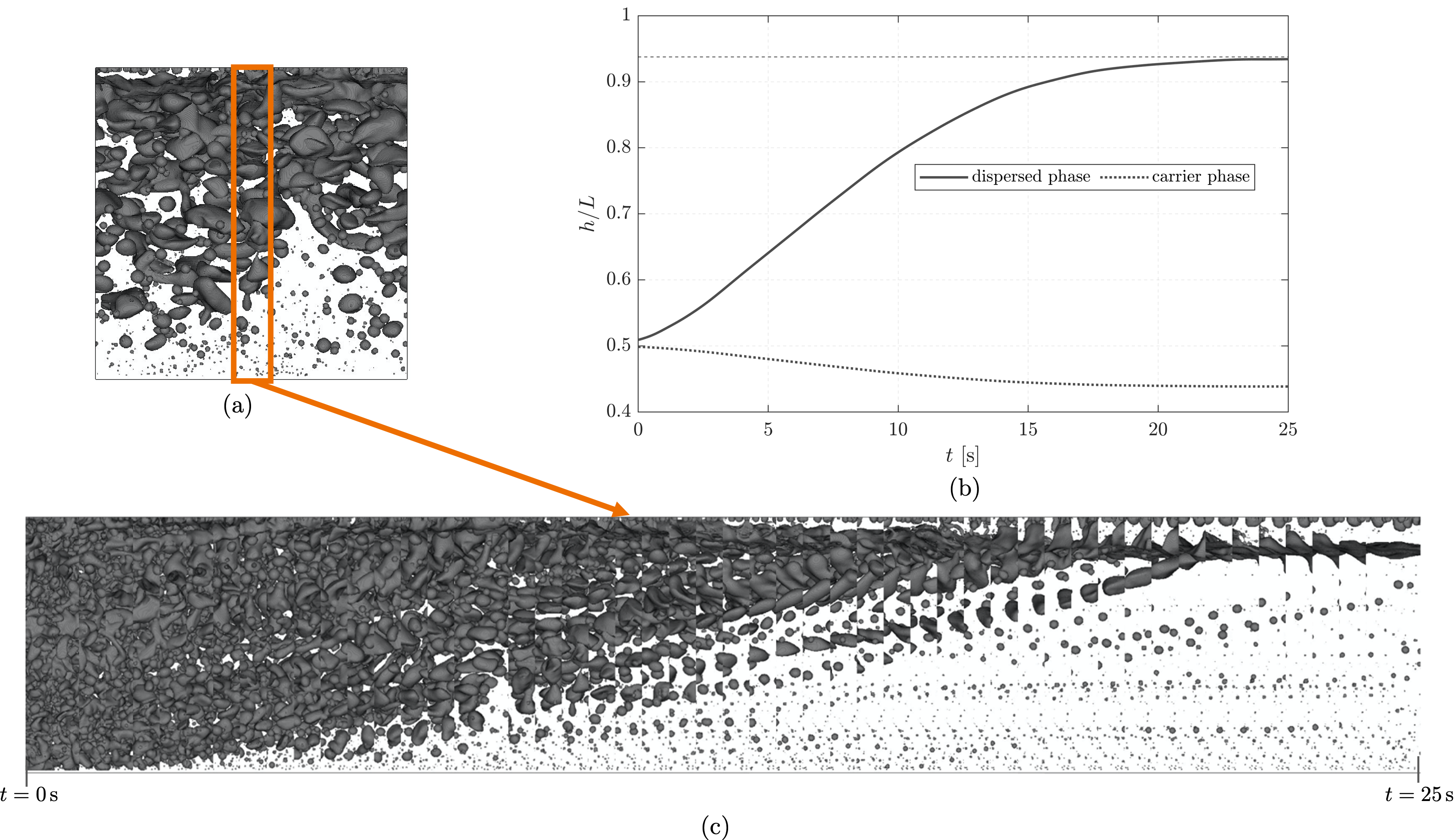}}
   \caption{Visualization of the segregation process. \new{(a) Front view on the emulsion, (b) recorded evolution of the center of mass of the dispersed phase, (c) time series of the orange section highlighted in (a) at a frame rate of 2 frames/s.}}
   \label{fig:ts_seg}
  \end{figure*}

  \new{Moreover, the time series in \cref{fig:SegG,fig:SegS} also illustrate the different droplet shapes during the segregation process. As expected, small droplets have a quasi-spherical shape due to the dominance of the surface tension forces, while larger droplets are rather ellipsoidal. \Cref{fig:SegS} shows the effect of the surface tension coefficient on the droplet shapes at comparable size. Comparing the last time steps of the \textit{Low}~$\sigma$ and the \textit{Baseline} case (\cref{fig:SegS}~(a-b,iv)), it can be seen that droplets of comparable size are more ellipsoidal for the \textit{Low}~$\sigma$ case. Regarding the droplet shape, there is an interplay between the surface tension force, which aims at a spherical shape, gravity, which affects the buoyancy force and also the hydrodynamic pressure inside the droplet, and the turbulent flow field. The well-known Grace-Diagram~\citep{grace1976shapes} allows for estimations of drop and bubble shapes as a function of the E\"{o}tv\"{o}s number, also known as Bond number, and the bubble Reynolds number. The E\"{o}tv\"{o}s/Bond number and its significance will be discussed in more detail in \cref{ss:Mechanisms}.
  }
 
  The segregation process under gravity can be characterized by the height of each phase and the interface area. In process engineering, mostly the height of the lighter phase or a coalescence or creaming interface is used to quantify the segregation, see e.g.~\citet{aleem2020experimental}. These quantities are optically easily accessible and are therefore commonly used. Note that the height refers to the position in the direction of gravitational acceleration. \newmk{In \cref{fig:ts_seg}, we have tried to adopt the experimental procedure for characterizing segregation to our simulation results and have specially post-processed them for this purpose.} Further, for a more accurate evaluation, we have monitored the height of the center of mass of each phase during our simulations. As can be seen in \cref{fig:ts_seg}~(b,c), the monitored height of the center of mass matches the optical impression. In a bounded domain, the changes of height are directly coupled by the correlation $\Delta h_c  = -\phi/(1-\phi)\Delta h_d$ (see also \cref{eq:h}), which can be seen in the visualization in \cref{fig:ts_seg}. In the following we only present the height of the center of mass of the dispersed phase, where the change is more evident. 
   \FloatBarrier
 \pagebreak
  \begin{figure*}[!tb]
    \subfigure[]{\includegraphics[trim={2cm 10cm 13cm 1cm},clip,width=0.49\linewidth]{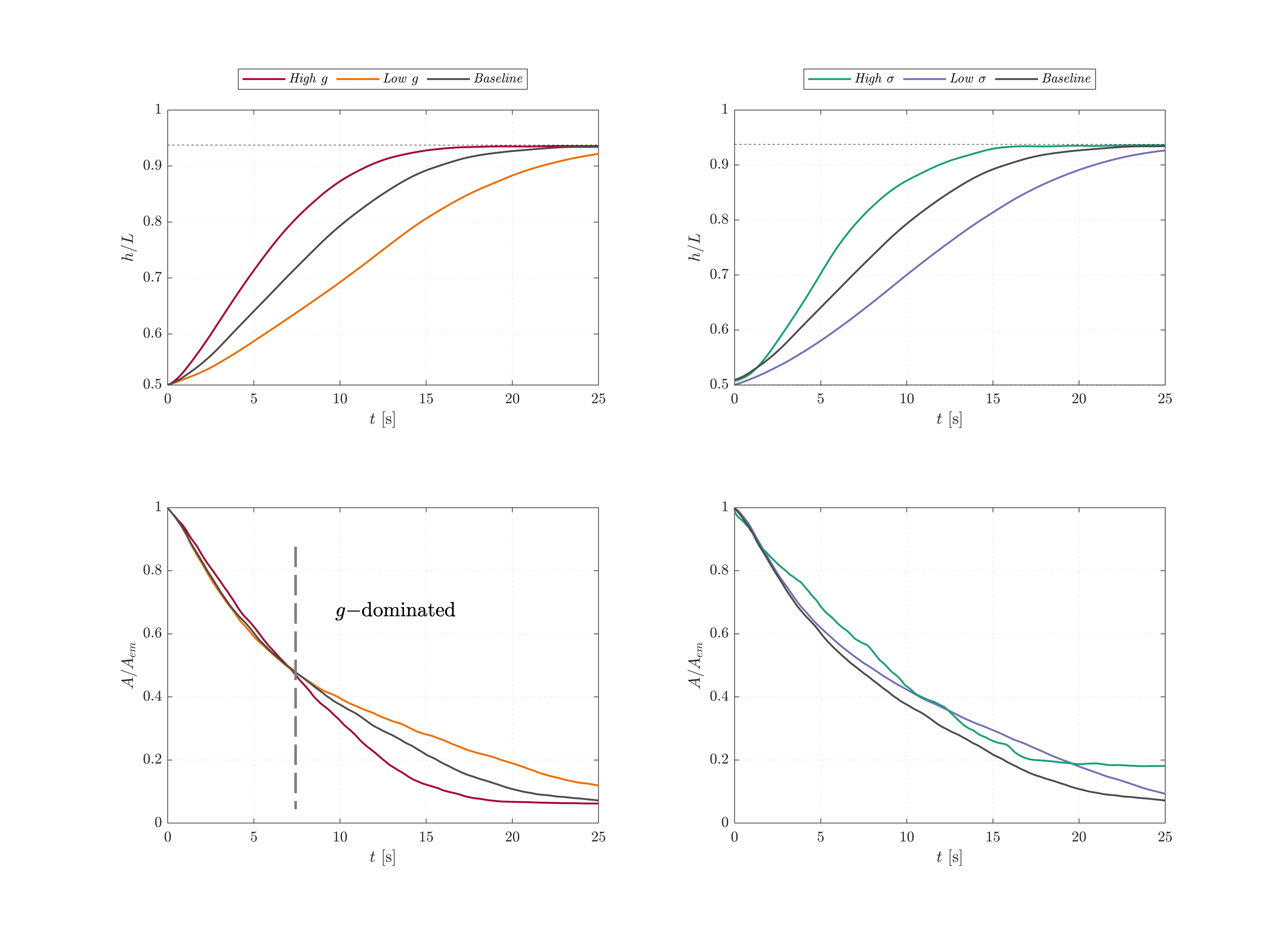}}
    \subfigure[]{\includegraphics[trim={14cm 10cm 1cm 1cm},clip,width=0.49\linewidth]{Fig5_PP}}
    \subfigure[]{\includegraphics[trim={2cm 1cm 13cm 10cm},clip,width=0.49\linewidth]{Fig5_PP}}
    \subfigure[]{\includegraphics[trim={14cm 1cm 1cm 10cm},clip,width=0.49\linewidth]{Fig5_PP}}
   \caption{Segregation progress measured by the height of the center of mass of the dispersed phase (a,b) and the interface area (c,d). The effect of different gravitational forces is shown in (a,c), whereas the effect of different surface tension coefficients is illustrated in (b,d).}
   \label{fig:HeightsComp}
  \end{figure*}

\Cref{fig:HeightsComp} visualizes the segregation progress measured by the height (a,b) and the interface area (c,d). The left column~(a,c) depicts the data for different gravity accelerations and the right column~(b,d) illustrates the progress for different surface tension coefficients. As discussed above, a stronger gravitational force promotes segregation and leads to a faster change in the heights of the center of mass. The surface tension coefficient also alters the segregation progress, since a higher surface tension coefficient accelerates the segregation measured of the height of the lighter phase. As can be seen in \cref{fig:SegS}, larger droplets are present for higher surface tension coefficients and the coalescence process is faster, which enhances the rise of the lighter phase. Additionally, the interface area can also be considered to characterize the segregation progress. For the present configuration, the recorded data is shown in \cref{fig:HeightsComp}~(c,d). It has to be noted that due to the upper and lower bounds in our configuration, structures must coalesce at a certain point in time, and thus the final coalescence process is clearly \newmk{dominated} by the gravitational acceleration. \Cref{fig:HeightsComp}~(c) reveals that after about $t=7\,\si{s}$ the segregation measured by the interface area is predominantly governed by the gravitational acceleration. Before that ($t<7\,\si{s}$), a smaller $g$ leads to a somewhat faster decay of the interface area, see \cref{fig:HeightsComp}~(c). We explain this by the fact that at a lower $g$ the structures remain at the same height a little longer and thus have more time to coalesce. For the evolution of the relative interface area $A/A_{em}$ at varying $\sigma$ (\cref{fig:HeightsComp}~(d)), no clear trend can be observed. It should be noted that the interface area of the emulsion $A_{em}$ is significantly larger for smaller surface tension coefficients, which biases the representation. The time derivative $\partial A/\partial t$, or more precisely that of the surface tension energy $\sigma \partial A/\partial t$, is more suitable for a comparison and will be analyzed in \cref{ss:ERR}. Further, the strongly fluctuating evolution at the \textit{High}~$\sigma$ case is related to the smaller number of droplets present in this configuration. 

  Overall, we conclude that for the present configuration, a clear correlation for the evolution of the interface can be expected only when there is no gravitational effect. The comparison of the evolution of the height and the interface area demonstrates that for the configurations studied, the height is more representative of the segregation progress and is therefore considered in the following. 

  \subsection{Driving mechanisms and dimensionless segregation number} 
  \label{ss:Mechanisms}

  Without energy input, emulsions are unstable and segregate over time as visualized in the previous section (\cref{ss:Results}). The segregation is driven by the energy release due to the minimization of the net potential energy and that of the surface energy. The change of potential energy of a two phase flow composed of a carrier and dispersed phase is given by
  \begin{equation}
   \Delta E_{pot} = (\rho_c V_c \Delta h_c +\rho_d V_d \Delta h_d) g\,.  
     \label{eq:EPot} 
  \end{equation}
  In a bounded domain, the change of heights is restricted to $V_c \Delta h_c + V_d\Delta h_d  = 0$. Using this and $V_d= \phi V$ and $V_c= (1-\phi) V$, the \newmk{relation}
  \begin{equation}
   \Delta h_c  = -\frac{\phi}{1-\phi}\Delta h_d  
   \label{eq:h} 
  \end{equation}
  is obtained. Thus, \cref{eq:EPot} can be simplified to 
   \begin{equation}
   \Delta E_{pot} = -\Delta \rho g V_d \Delta h_d 
  \label{eq:EPot2} 
  \end{equation} 
  with $\Delta \rho=\rho_c-\rho_d$, which is in the considered configuration positive since $\rho_c > \rho_d$. Consequently, the rise of the lighter phase (here dispersed phase) releases energy ($\Delta E_{pot}<0$). 

  The change of surface energy is given by 
  \begin{equation}
   \Delta E_{\sigma} = \sigma \Delta A.
  \label{eq:ES} 
  \end{equation}
  Breakup leads to an increase of the interface area ($\Delta A>0$) and requires energy input, while coalescence leads to a reduction of the interface area ($\Delta A<0$), thus releasing energy ($\Delta E_{\sigma}<0$). 

  To identify the dominant mechanism promoting the segregation, we propose a non-dimensional energy release ratio $\Psi$ of these two driving mechanisms 
  \begin{equation}
    \Psi = \frac{\Delta E_{pot}}{\Delta E_{\sigma}}=\frac{ -\Delta \rho g V_d \Delta h_d} {\sigma \Delta A}.
  \label{eq:ERR1} 
  \end{equation}
  The correlation above depends on several case specific quantities as $V_d$, $\Delta h_d$ and $\Delta A$, where especially the latter is often a-priori unknown. In the following we aim to derive a dimensionless correlation, which depends only on the fluid properties $\Delta \rho$ and $\sigma$, the gravitational acceleration $g$ and two characteristic length scales - $d$ for a representative droplet size and $H$ for a representative length scale of the segregation progress. To this end, we substitute the change of the interface area $\Delta A= A_\infty - A_{em}$, where $A_\infty$ is the interface area at fully segregated state and $A_{em}$ is the interface area of the emulsion at stationary state, by $\Delta A\approx -A_{em}$, which is valid for $A_\infty << A_{em}$. Further, $A_{em}$ can be expressed with the Sauter mean diameter $d_{32}=6 V_d /A_{em}$, for which constant relations to the Hinze scale have been reported $d_{32} \propto d_H$ \citep{pacek1998sauter,Yi:2020dn,boxall2012droplet}. This allows for the approximation
  \begin{equation}
    \Delta A \approx -6 V_d/d_{32}. 
  \label{eq:DeltaA} 
  \end{equation}
  Additionally substituting $\Delta h_d$ with a representative length scale $H$ gives
    \begin{equation}
    \Psi \approx \frac{1}{6}\frac{\Delta \rho g d_{32} H} {\sigma }.
  \label{eq:ERR2} 
  \end{equation}
  Using this correlation, a dimensionless segregation number $Seg$ can be defined as
  \begin{equation}
    Seg =\frac{\Delta \rho g d H} {\sigma },
  \label{eq:ERR2} 
  \end{equation}
  where the representative length scale for the segregation process $H$ is the height the lighter (here dispersed) phase rises on average. In a bounded domain with length $L$, $H$ can be calculated using $H = (1 - 0.5\phi ) L - 0.5 L = 0.5(1-\phi) L$, which here is $H =7/8\pi$. $d$ stands for a representative diameter of the emulsion, which can be approximated with $d_{32}$ or $d_H$. It is worth noting that the dimensionless segregation number $Seg$ closely resembles the dimensionless Bond number $Bo$, also known as E\"otv\"os number, with
  \begin{equation}
   Bo=\frac{\Delta \rho g R^2} {\sigma },
  \label{eq:Bo} 
  \end{equation}
  for rising/falling bubbles or droplets with radius $R$. However, $Bo$ has a different physical relevance since it describes the ratio of body forces to surface forces and characterizes the bubble/droplet shape and the tendency for a breakup. \new{$R$ in \cref{eq:Bo} can be substituted by a different characteristic length scale of the bubble/droplet.} \newnew{Because of their different physical meanings, $Seg$ and $Bo$ use different length scales. The length scale for $Bo$ is associated with the bubble/droplet, while for $Seg$ the product of a length scale associated with the bubble/droplet (denoted here by $d$) and one associated with the segregation process (denoted here by $H$) is used. Note that a Bond number $Bo$ with a characteristic length scale of $\sqrt{d H}$, which is physically difficult to motivate, leads to the same expression as $Seg$.}

  \Cref{tab:CasesSeg} contains the segregation number $Seg$ determined with $d_H$ together with the dimensionless energy release ratio $\Psi$ evaluated using the changes between $t = 0\,\si{s}$ and $t = 25\,\si{s}$. A higher $Seg$ (or higher $\Psi$) indicates that the release of potential energy dominates for segregation, while a lower number implies a more important role of the surface tension energy release. It should be noted that the segregation number $Seg$ (or $\Psi$) only indicates the ratio of the two energy releases and does not provide any information about the timescale of the segregation process, which is analyzed in \cref{ss:u}. 
  
  The ratio $Seg/\Psi$ is also included in \cref{tab:CasesSeg}. For a variation of $g$, the ratio is constant confirming the validity of the above made approximations. When $\sigma$ is varied, the ratio is approximately the same, but the values scatter. We conjecture that this is due to the approximation used for $\Delta A$ (\cref{eq:DeltaA}).

  \begin{figure*}[!tb]
    \subfigure[]{\includegraphics[trim={2cm 0cm 17cm 0cm},clip,width=0.49\linewidth]{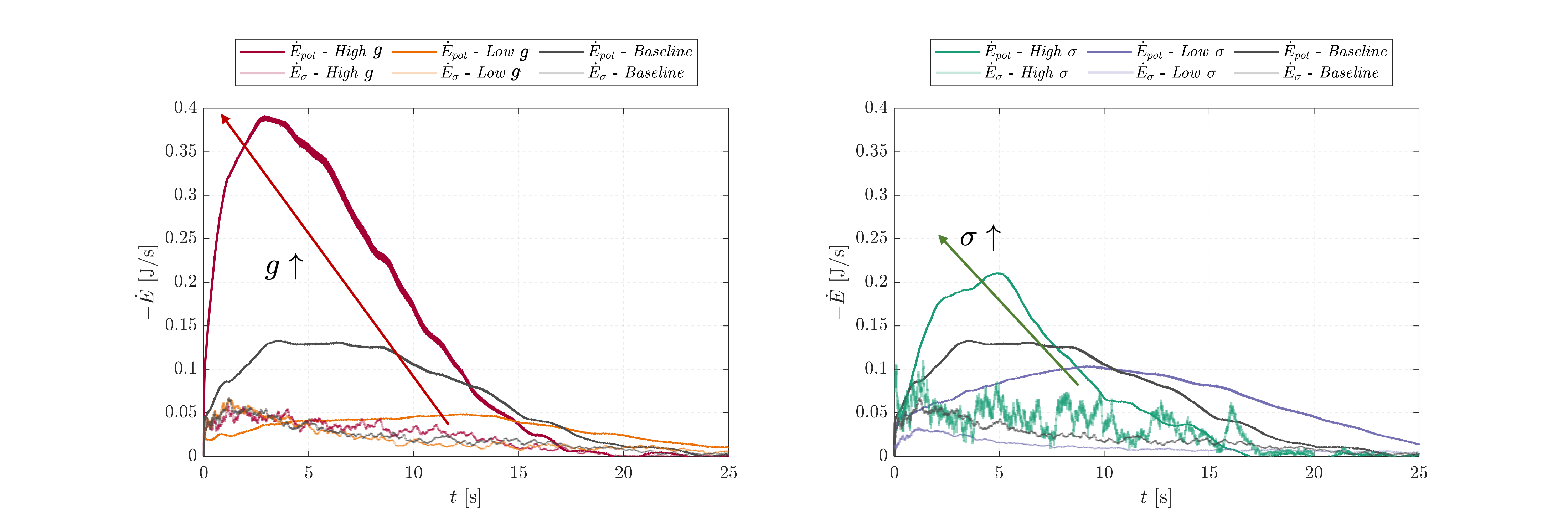}}
    \subfigure[]{\includegraphics[trim={17cm 0cm 2cm 0cm},clip,width=0.49\linewidth]{Fig6_PP_v2}}
   \caption{Energy release rates due to the reduction of the net potential height $\dot{E}_{pot}$ and the reduction of the interface area $\dot{E}_{\sigma}$ for varying $g$ in (a) and varying $\sigma$ in (b). } 
  \label{fig:ERates}
  \end{figure*}

  \begin{figure*}[!tb]
    \subfigure[]{\includegraphics[trim={2cm 0cm 15cm 0.5cm},clip,width=0.49\linewidth]{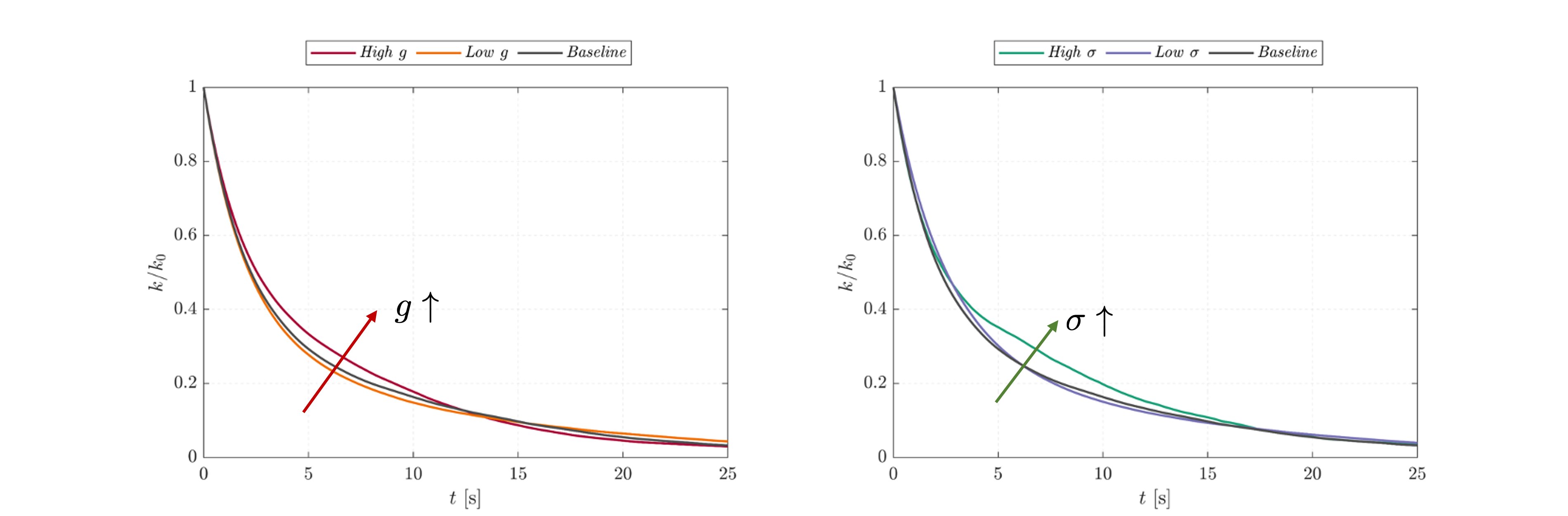}}
    \subfigure[]{\includegraphics[trim={15cm 0cm 2cm 0.5cm},clip,width=0.49\linewidth]{Fig7_PP_v2}}
   \caption{Decay of the normalized kinetic energy in different configurations. (a) for varying $g$, (b) for varying $\sigma$.}
   \label{fig:Ekin}
  \end{figure*}

  \subsection{Energy release rates and decay of turbulent kinetic energy} 
  \label{ss:ERR}

  For the actual segregation, not only the amount of energy release is crucial, but also the time over which it is released. To this end, we propose to evaluate energy release rates. Using \cref{eq:EPot2,eq:ES}, the energy release rates are 
  \begin{equation}
   \dot{E}_{pot} = \Delta \rho V_d \dot{h}_d g  \quad  \dot{E}_{\sigma} = \sigma \dot{A}. 
  \label{eq:Rates} 
  \end{equation}
  As discussed above, during the segregation progress both quantities are negative and therewith drive this process. \Cref{fig:ERates} visualizes the energy release rates. For the variation of the gravitational acceleration (\cref{fig:ERates}~(a)), the energy release for the potential energy $\dot{E}_{pot}$ clearly increases with increasing $g$, while the energy release due to the reduction of the surface energy $\dot{E}_{\sigma}$ is similar \newmk{for  the three $g$-variations, see \cref{fig:ERates}~(a).} The difference between $\dot{E}_{pot}$ at different $g$ is not only caused by the different $g$ values, but also the resulting different $\dot{h}$, amplifying the differences in the energy release. The variation of the surface tension coefficient (see \cref{fig:ERates}~(b)) does not reveal such a clear trend. As expected, the energy release from the surface tension term $\dot{E}_{\sigma}$ increases with increasing $\sigma$, however, the value of $\sigma$ also affects the release of the potential energy $\dot{E}_{pot}$. At a smaller $\sigma$, the dispersed phase remains in a more disturbed state, i.e., smaller droplets, for a longer time. This leads to a greater number of droplet interactions and mutual hindering effects, limiting the rise of the lighter phase and thus the release of potential energy. Conversely, a higher $\sigma$ means fewer droplets as well as faster coalescence and, consequently, less restriction on the release of potential energy. Hence, the release of potential energy appears to scale proportional to the droplet size in the emulsions, see also \cref{fig:ERates}~(b). 
  
  Further, the release of potential and surface energy affects the decay of the turbulent kinetic energy, which is depicted in \cref{fig:Ekin}. \Cref{fig:Ekin}~(a) shows that a higher $g$ results in higher energy release due to the decreasing potential energy, which is transformed in kinetic energy and retards the decay of the latter. The variation of $\sigma$ also affects the decay of turbulent kinetic energy (\cref{fig:Ekin}~(b)). However, due to the complex interplay of droplet size distributions and the release of gravitational energy, no clear trend is observed. For a detailed study of the effect of varying $\sigma$ on the decay of turbulent kinetic energy in decaying turbulence without gravitational force, we refer the reader to \citet{Dodd:2016eo}.
   \FloatBarrier

  \subsection{Time scale of the segregation process}
  \label{ss:u}

   For practical applications, the duration of the segregation process and the effects of parameter changes on this duration are of particular interest. Therefore, we attempt to derive a characteristic time scale for segregation progress. For this purpose, we consider the time evolution of the height of the lighter phase and its time derivative which represents an average rising velocity. 

     \begin{figure*}
    \includegraphics[width=0.99\linewidth]{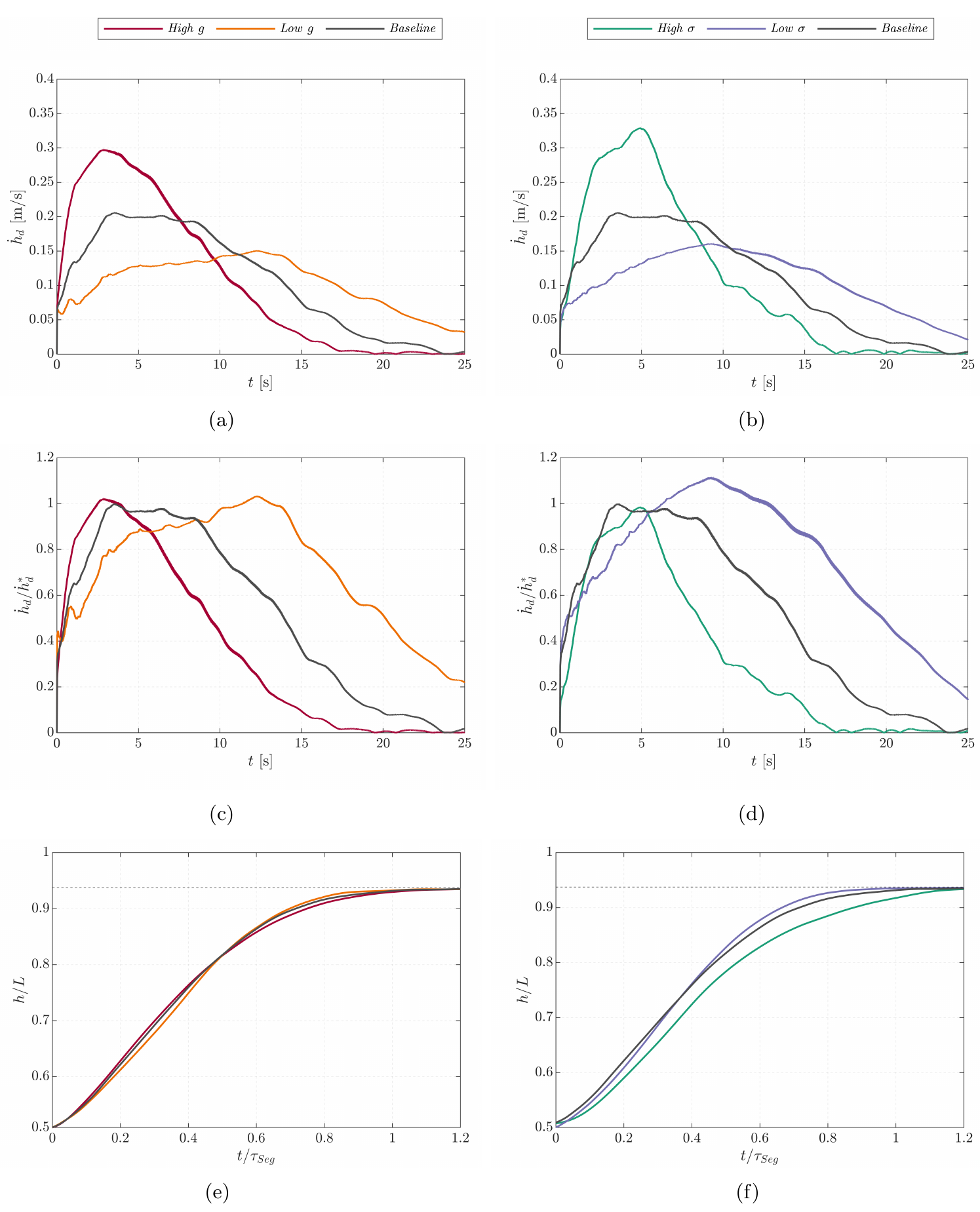}
   \caption{Temporal derivative of the center of mass of the dispersed phase $\dot{h}_d$ (a,b), normalized temporal derivative $\dot{h}_d/\dot{h}_d^*$ (c,d) and temporal evolution of \newmk{$h/L$} scaled by the proposed timescale (e,f). The left column (a,c,e) shows these quantities for different gravitational accelerations $g$ and the right column (b,d,f) illustrates them for different surface tension coefficients $\sigma$.
   \newline\newline\newline\newline\newline\newline\newline\newline}
   \label{fig:Velocities}
  \end{figure*}

  \Cref{fig:Velocities}~(a,b) illustrates the temporal derivative of the height of the center of mass of the dispersed phase $\dot{h}_d$. As discussed in the previous subsection, in case of a variation of $g$, $\dot{h}_d$ clearly increases for an increasing $g$, see \cref{fig:Velocities}~(a). The droplet size distribution (associated with $\sigma$) alters the release rate of potential energy and thus $\dot{h}_d$. A more dispersed emulsion, characterized by a smaller $d_H$, has a higher hindering effect and limits the release of the potential energy, whereas a less dispersed emulsion, characterized by a higher $d_H$, allows for a higher energy release of potential energy. Consequently, $\dot{h}_d$ increases with increasing $\sigma$ (increasing $d_H$), see \cref{fig:Velocities}~(b).

  The average rising velocity of a single droplet due to gravitational acceleration, neglecting friction forces, is given by 
    \begin{equation}
    U_\mathrm{g} = \frac{H}{t}= \sqrt{\frac{1}{2}\frac{\Delta \rho } {\rho_d}g H},
    \label{eq:Ug_0} 
  \end{equation}
  where $H$ stands for the height, which the dispersed phase has risen. The detailed derivation of this \newmk{relation} is provided in the appendix. This velocity, of course, significantly overestimates the average rising velocity of the considered configuration as friction forces and droplet--droplet interactions are neglected. However, the gravity-based velocity (\cref{eq:Ug_0}), together with the observations described above, motivate the formulation
  \begin{equation}
    \dot{h}_d  \propto U_g  \xi(d)\,,  
    \label{eq:v} 
  \end{equation}
  where $\dot{h}_d$ is assumed to be proportional to a gravity based velocity $U_g$ and a factor $\xi$ depending on the droplet size distribution of the emulsion. The proposed \newmk{relation} is a simplification for the configurations considered here. For other configurations, additional effects of other parameters, such as viscosity, would also have to be incorporated. We approximate $d$ with the Hinze scale $d_H$ and propose 
  \begin{equation}
    \xi = (d_H/d_{ref})^\gamma\, .
    \label{eq:v2} 
  \end{equation}%
  For the considered configuration, we have fitted $\gamma =  0.5$ and for simplicity taken $d_{ref} = d_{H\mathrm{BL}}$. This results in the correlation
  \begin{equation}
    \dot{h}_d^*  = c_p U_g \xi( d) =  c_p \sqrt{\frac{1}{2}\frac{\Delta \rho } {\rho_d}g H} \sqrt{\frac{d_H}{ d_{ref}}},  
    \label{eq:v3} 
  \end{equation}
  where $c_p$ denotes a \newmk{proportionality} factor depending on the choice of $d_{ref}$. For $d_{ref} = d_{H\mathrm{BL}}$, $c_p$ is fitted to $c_p = 0.25$. The measured velocities $\dot{h}_d$ normalized by $\dot{h}_d^*$ (\cref{eq:v3}) are plotted in \cref{fig:Velocities}~(c,d). \newmk{In all cases, the normalized velocities have their maximum at about 1, which confirms the validity of the approximations made.}

  Further, the derived correlation allows for an estimation of a timescale for the segregation. Using \cref{eq:v,eq:v2} the following is obtained: 
  \begin{equation}
      \tau_\mathrm{Seg} \propto \frac{H}{U_{g} (d_H/d_{ref})^{\gamma}} \,.   
    \label{eq:t_g2} 
  \end{equation}
  Using additionally the definition of $U_\mathrm{g}$ (\cref{eq:Ug_0}) results in 
  \begin{equation}
     \tau_\mathrm{Seg} \propto \sqrt{\frac{\rho_d H}{\Delta \rho g}}\left(\frac{d_{ref}}{d_H}\right)^{\gamma}\,.  
  \label{eq:t_g3} 
  \end{equation}
 Thus, for a variation of only $g$, the timescale is proportional to $\tau \propto 1/\sqrt{g}$ and for a variation of only the droplet size distribution ($\sigma$), the timescale is proportional to $\tau \propto 1/{{d_H}^{\gamma}}$. 

 For the considered configuration, the timescale of the segregation can be explicitly calculated as 
   \begin{equation}
     \tau_\mathrm{Seg} =  \frac{H}{c_p U_{g} (d_H/d_{ref})^{\gamma}}.  
  \label{eq:t_g4} 
  \end{equation}
 Using $\gamma = 0.5$, $c_p = 0.25$, and \cref{eq:Ug} gives 
    \begin{equation}
     \tau_\mathrm{Seg} =  2^{5/2} \sqrt{\frac{\rho_d H}{\Delta \rho g}} \sqrt{\frac{d_{ref}}{d_H}},   
  \label{eq:t_g5} 
  \end{equation}
  where $H = 7/8\pi$. \Cref{fig:Velocities}~(e,f) shows the height of the dispersed phase $h$ plotted over the time normalized by the characteristic timescale. It can be seen that for both variations, the segregation measured by height is completed at $\tau_\mathrm{Seg}$. Moreover, it is noteworthy that the scaled temporal evolution of the heights almost coincides to one line when $g$ is varied, see \Cref{fig:Velocities}~(e). 

\section{Conclusions}
  \label{s:Conclusions} 
  
  In this work, we \new{have} numerically \new{studied} the segregation of turbulent emulsions under different gravitational accelerations $g$ and with different droplet size distributions obtained by altering the surface tension coefficient $\sigma$. To this end, we first generated turbulent emulsions in a linearly forced HIT and then turned off the forcing and activated the gravitational acceleration. \new{This approach enabled us to study the segregation process using well-defined initial conditions.} \new{To our knowledge, this work represents the first numerical investigation of the gravity-driven segregation process. With it, we extend previous numerical studies focusing on emulsification or emulsions at the statistically stationary state. Moreover, the time-resolved, three-dimensional visualization of the segregation progress obtained by our DNS studies supplements existing experimental studies on segregation.} We have approached this topic from the thermodynamic perspective of energy releases, adding an important complementary perspective to this physical process.

  Segregation can be quantified by the height of each phase and the interface area. In the present study, we have primarily analyzed the temporal evolution of the height of the dispersed phase. Moreover, we have addressed the energy release of the two central processes, namely the rise of the lighter phase (release of potential energy) and coalescence (release of surface energy). Based on our observations, we have defined a dimensionless segregation number $Seg$ that characterizes the ratio of potential energy release to surface energy release, allowing for an identification of the dominating process. In addition, we evaluated and compared the energy release rates. Our simulation results show that a smaller droplet size, i.e., smaller $\sigma$, hinders and limits the release of potential energy. 

  Finally, we have derived a correlation to estimate the average rising velocity of the lighter phase, which also allows for the derivation of a characteristic timescale. We found that the average rising velocity is a fraction of a gravity based velocity and depends on the size of the droplets in the emulsion. Scaling of the velocities and the time with the empirically derived correlation showed good agreement. 

  The presented work can be considered as a first important step towards the numerical assessment of emulsion segregation. Subject of current investigations is the evaluation of droplet size distributions during the segregation process. Therefore, in order to obtain statistically reliable data, a multitude of identical segregation processes have to be simulated simultaneously. Furthermore, in future studies we plan to consider configurations with varying density differences between the dispersed and carrier phase to assess the effect of the density difference on the segregation and its time scale. 

\section*{Declaration of Competing Interest}

  The authors declare that they have no known competing financial interests or personal relationships that could have appeared to influence the work reported in this paper.

\section*{Acknowledgment}

  This project received funding by \textit{dtec.bw} - Digitalization and Technology Research Center of the Bundeswehr - under the project \textit{MORE}, which is gratefully acknowledged. Further, the authors thank the Gauss Centre for Supercomputing e.V. (www.gauss-centre.eu) for funding this project by providing computing time on the GCS Supercomputer SuperMUC-NG at Leibniz Supercomputing Centre (www.lrz.de).

\section*{Data Availability}

  The data that support the findings of this study are available from the corresponding author upon reasonable request.
\FloatBarrier
\section*{Appendix}
  
  \subsection*{Gravity based velocity derived from the force balance on a droplet}
  \label{ss:ug}

  A gravity-based characteristic velocity $U_\mathrm{g}$ can be derived based on the force balance on a droplet with mass $m$ ($m=\rho_d V_{droplet}$). Neglecting friction forces and other losses the force balance reads
  \begin{equation}
    m_d \ddot{x}= F_b-F_g, 
  \end{equation}
  where $F_b$ denotes the buoyancy force with $F_b=\rho_c V_{droplet} g$ and $F_g$ denotes the gravitation force with $F_g=\rho_d V_{droplet} g$. This leads to the following acceleration: 
  \begin{equation}
      \ddot{x}= \frac{\Delta \rho}{\rho_d} g.  
    \label{eq:xDotDot} 
  \end{equation}
  Integrating twice in time and using the initial conditions $\dot{x}=0,\,x=0$ gives
  \begin{equation}
    x = \frac{\Delta \rho}{\rho_d} g \frac{1}{2}t^2. 
    \label{eq:x} 
  \end{equation}
  The distance over a time $t$ is set to $x=H$ and $t$ can be expressed as  
  \begin{equation}
    t = \sqrt{2 \frac{\rho_d}{\Delta \rho}\frac{H}{g}}. 
    \label{eq:t} 
  \end{equation}
  Thus, the average velocity resulting from gravitation over a distance $H$ can be determined with
  \begin{equation}
    U_\mathrm{g} = \frac{H}{t}= \sqrt{\frac{1}{2}\frac{\Delta \rho } {\rho_d}g H}.
    \label{eq:Ug} 
  \end{equation}
  This \newmk{relation} can alternatively be derived by evaluating the velocity based on the kinetic energy equivalent of the release of potential energy, as shown below.

  \subsection*{Gravity based velocity derived from the energy release of the potential energy}

    The gravity based characteristic velocity (\cref{eq:Ug}) can also be derived by evaluating the velocity based on the kinetic energy equivalent of the  release of potential energy. For consistency, we also consider the kinetic energy in the carrier phase. This gives the following balance  
    \begin{equation}
     \frac{1}{2} \rho_d V_d U_{g,max}^2+\frac{1}{2} \rho_c V_c U_{c,g,max}^2 =   - \Delta E_{pot}, 
     \label{eq:Ebil1} 
    \end{equation} 
    where $U_{g,max}$ refers to the maximum velocity in the dispersed phase to be consistent with the nomenclature used so far and $U_{c,g,max}$ to that in the carrier phase. Using \cref{eq:h}, we can recast the left part to
    \begin{equation}
      \frac{1}{2}\rho_d V_d U_{g,max}^2(1+\frac{\rho_c}{\rho_d}\frac{\phi}{1-\phi}) =  - \Delta E_{pot}.
      \label{eq:Ebil2} 
    \end{equation} 
    We simplify this expression with $\beta = \frac{\rho_c}{\rho_d}\frac{\phi}{1-\phi}$. Note that for small void fractions and density ratios close to 1 this expression vanishes ($\beta\to 0$). Inserting the expression for the release of potential energy from \Cref{eq:EPot2} yields
    \begin{equation}
      \frac{1}{2}\rho_d V_d U_{g,max}^2(1+\beta) = \Delta \rho V_d g \Delta h_d.
      \label{eq:Ebil3} 
    \end{equation} 
    Based on this correlation, we can derive a maximum velocity associated with the energy release of the potential energy that reads
    \begin{equation}
      U_{g,max} =  \sqrt{\frac{2}{1+\beta} \frac{\Delta \rho }{\rho_d} g \Delta h_d}.
      \label{eq:Udmaxg} 
    \end{equation}
    Assuming that the initial velocity is zero and the acceleration is linear, the average velocity can be approximated by $U_\mathrm{g} \approx \frac{1}{2} U_{g,max}$. Further, we substitute $\Delta h_d$ with $H$ and obtain 
    \begin{equation}
      U_{g} =  \sqrt{\frac{1}{2(1+\beta)} \frac{\Delta \rho }{\rho_d} g H}.  
      \label{eq:Ug2} 
    \end{equation}
    For $\beta=0$ this expression is identical to \cref{eq:Ug}. In the considered configurations we have $\beta=0.1587$ which corresponds to a prefactor of $\sqrt{1/(1+\beta)}=0.9289$ for $U_{g}$ and has been neglected in the evaluations shown in the paper for the sake of simplicity.

\FloatBarrier

\section*{References}
\bibliography{pof}
\end{document}